\documentclass[sigconf, nonacm]{acmart}

\newcommand\vldbdoi{XX.XX/XXX.XX}

\newcommand\vldbvolume{14}
\newcommand\vldbissue{1}

\newcommand\vldbavailabilityurl{URL_TO_YOUR_ARTIFACTS}
\newcommand\vldbpagestyle{plain} 
\usepackage{xspace}
\newcommand{\system}{{\sc DataMagic}\xspace}
\newcommand{\dvspec}{{\sc DVSpec}\xspace}

\usepackage{enumitem}

\newcommand{\stab}{\vspace{1.2ex}\noindent}
\newcommand{\stitle}[1]{\stab\noindent{\textbf{#1}}}

\begin{document}
\title{DataMagic: Transforming Tabular Data into Data Insight Video}

\author{Yupeng Xie}
\affiliation{%
  \institution{HKUST (GZ)}
}

\author{Chen Ma}
\affiliation{%
  \institution{China Unicom}
}

\author{Zhenyang Wang}
\affiliation{%
  \institution{HKUST (GZ)}
}

\author{Liangwei Wang}
\affiliation{%
  \institution{HKUST (GZ)}
}

\author{Jiayi Zhu}
\affiliation{%
  \institution{HKUST (GZ)}
}

\author{Chuxuan Zeng}
\affiliation{%
  \institution{China Unicom}
}

\author{Zhouan Shen}
\affiliation{%
  \institution{HKUST (GZ)}
}

\author{Boyan Li}
\affiliation{%
  \institution{HKUST (GZ)}
}

\author{Yuyu Luo}
\authornote{Yuyu Luo is the corresponding author.}
\affiliation{%
  \institution{HKUST (GZ)}
}

\begin{abstract}
Data videos integrate dynamic charts, voice narration, and synchronized animations to communicate data insights as temporal narratives, making them an effective medium for improving data consumption efficiency in the data management lifecycle. 
{However, producing high-quality data videos requires expertise spanning data analysis, narrative design, and video production.}
Existing approaches fall short: static visualization tools (e.g., BI dashboards) lack narrative logic and animation; authoring tools require users to pre-prepare visualizations rather than working from raw data; pixel-level video generation models cannot guarantee data fidelity or provenance. We demonstrate \system, an end-to-end interactive system that transforms raw tabular data and natural language queries into narrative data-insight videos. To ensure data fidelity, \system introduces the declarative specification \dvspec, which binds visual and animation elements to underlying data fields through data-driven semantic references. To address the combinatorial explosion of the design space, \system adopts a Generate-then-Orchestrate multi-agent architecture that generates candidate scenes in parallel and then optimizes narrative coherence through global orchestration. Leveraging \dvspec's decoupling of logic and rendering, the system further supports three interaction modes and structured provenance-based data Q\&A, transforming one-way videos into explorable interactive data interfaces. Evaluation on 109 real-world samples validates the effectiveness of the \system.
\end{abstract}

\maketitle

\pagestyle{\vldbpagestyle}
\begingroup
\renewcommand\thefootnote{}\footnote{\noindent
This work is licensed under the Creative Commons BY-NC-ND 4.0 International License. Visit \url{https://creativecommons.org/licenses/by-nc-nd/4.0/} to view a copy of this license. For any use beyond those covered by this license, obtain permission by emailing \href{mailto:info@vldb.org}{info@vldb.org}. Copyright is held by the owner/author(s). Publication rights licensed to the VLDB Endowment. \\
\raggedright Proceedings of the VLDB Endowment, Vol. \vldbvolume, No. \vldbissue\ %
ISSN 2150-8097. \\
\href{https://doi.org/\vldbdoi}{doi:\vldbdoi} \\
}\addtocounter{footnote}{-1}\endgroup


\vspace{-2em}
\section{Introduction}

Data videos integrate dynamic charts, voice narration, and synchronized animations to communicate data insights as temporal narratives, attracting growing attention from both academia and industry~\cite{amini2015understanding, shen2025reflecting, wang2026tabletale}. However, producing high-quality data videos requires cross-domain expertise spanning data analysis, narrative design, and video production~\cite{shen2025reflecting}, and existing approaches all fall short of end-to-end automation from raw data to data video~\cite{seedb, zql, li2026deepeye, tang2026vividoc, tang2026demonstrating}.

Static visualization tools (e.g., BI dashboards, HAIChart~\cite{xie2024haichart}, DeepEye~\cite{luo2018deepeye}) output static charts, lacking narrative logic and animation. Authoring tools (e.g., Data Playwright~\cite{shen2025dataplaywright}) focus on adding animation effects to existing charts, without extracting insights from raw data or handling multi-scene narrative orchestration. Pixel-level video generation models (e.g., Sora~\cite{liu2024sora}) can synthesize videos, but their black-box nature frequently produces numerical hallucinations and cannot map visual elements back to underlying data records.

Our key observation is that effective data videos are fundamentally structured narratives rather than simple assemblies of visual elements~\cite{shen2025reflecting}. This motivates us to model end-to-end generation as a hierarchical content orchestration problem. Two core challenges emerge: (1)~how to design a structured intermediate representation that precisely describes heterogeneous components and their temporal relations while ensuring data fidelity and provenance; (2)~how to efficiently search the vast design space for solutions that balance local scene quality and global narrative coherence.

To this end, we demonstrate \system\footnote{\textcolor{blue}{\textbf{\system: \url{https://datamagic-home.github.io/}}}}, an end-to-end interactive system from raw tabular data to narrative data videos. For challenge~(1), we introduce \dvspec (Data Video Specification), a declarative specification that binds visual and animation elements to underlying data fields through data-driven semantic references and narration-index triggering, ensuring full provenance. For challenge~(2), \system adopts a Generate-then-Orchestrate multi-agent architecture~\cite{zhu2025survey} that generates diverse candidate scenes in parallel and then applies global orchestration to optimize scene selection, ordering, and narrative coherence. Leveraging \dvspec's decoupling of logic and rendering, the system further supports three interaction modes and structured provenance-based data Q\&A, enabling users to refine videos and explore data directly.

The main contributions are as follows: (1)~We demonstrate \system, introducing the declarative specification \dvspec and the Generate-then-Orchestrate multi-agent strategy, which mechanistically ensure data fidelity and provenance. (2)~We design three progressive demonstration scenarios covering automated generation, multimodal editing, and provenance-based data Q\&A, and validate system effectiveness on 109 real-world samples.

\begin{figure*}[t]
    \centering
    \includegraphics[width=0.95\textwidth]{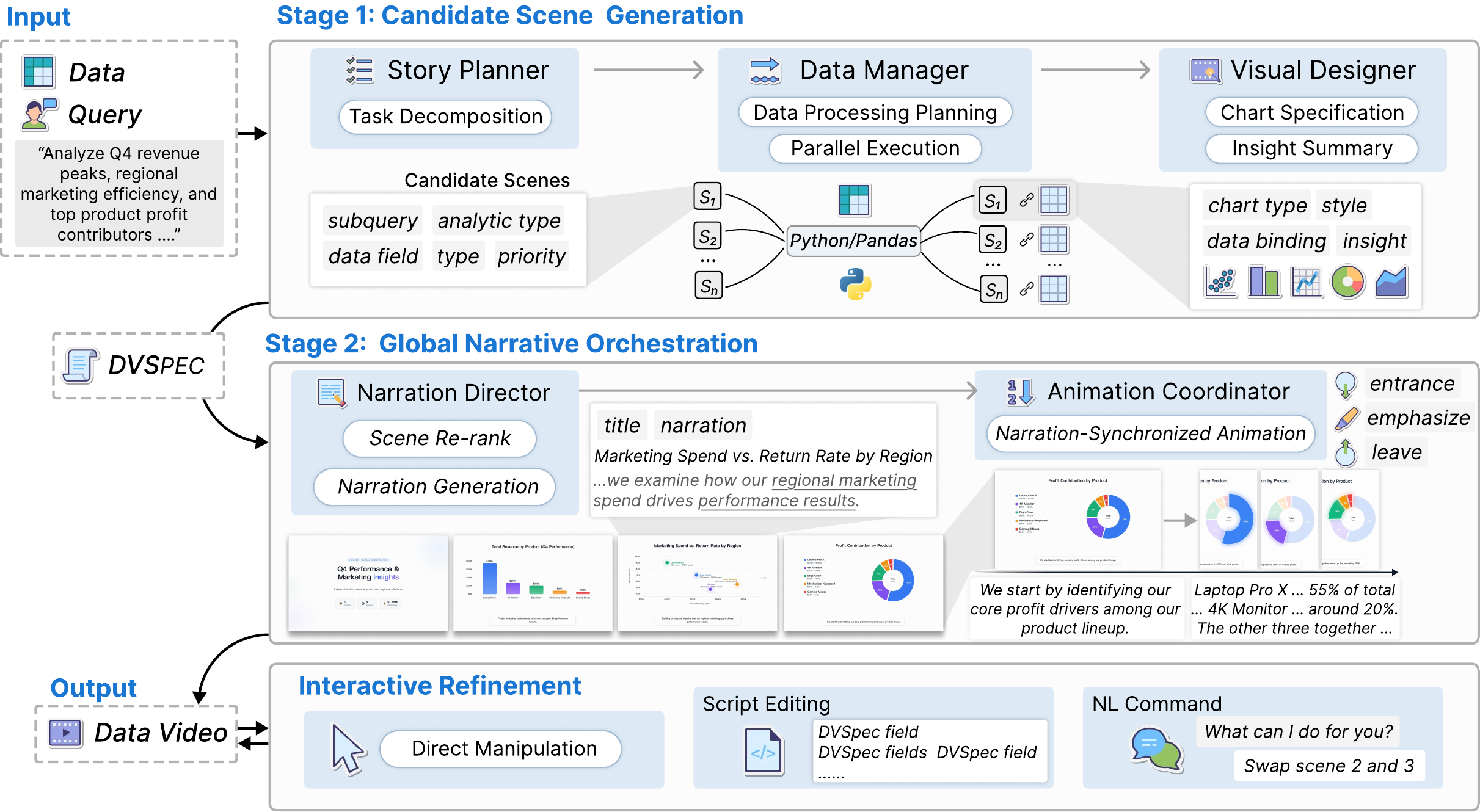}
    \caption{System architecture of \system. 
    }
    \label{fig:architecture}
  \end{figure*}

\section{System Architecture}

As shown in Figure~\ref{fig:architecture}, \system takes raw tabular datasets and natural language queries as input, processes them through a multi-agent engine to produce a \dvspec configuration, and compiles it into a complete narrative video. The core design revolves around three components: \dvspec declarative specification (Section~\ref{sec:dvspec}), the multi-agent generation pipeline (Section~\ref{sec:pipeline}), and provenance-based interaction and exploration (Section~\ref{sec:editing}).

\subsection{\dvspec: Declarative Data Video Specification}
\label{sec:dvspec}

Existing declarative specifications (e.g., Vega-Lite~\cite{satyanarayan2017vegalite}, Canis~\cite{ge2020canis}, ChartMark~\cite{chen2025chartmark}) perform well for static charts or single-chart annotations and animations, but have not been extended to the unified description of cross-modal content and temporal coordination required for multi-scene data videos{~\cite{jade, dig}}. To fill this gap, we design \dvspec (Data Video Specification), a declarative specification that fully decouples logical description from rendering implementation, serving as a structured intermediate representation between the multi-agent generation engine and the rendering engine: the generation stage writes analysis results to \dvspec, while interactive editing changes are also mapped to local \dvspec updates. \dvspec itself is rendering-library agnostic; the current system renders charts using D3.js and synthesizes videos with Remotion.

\begin{figure}[t]
  \centering
  \includegraphics[width=1\linewidth]{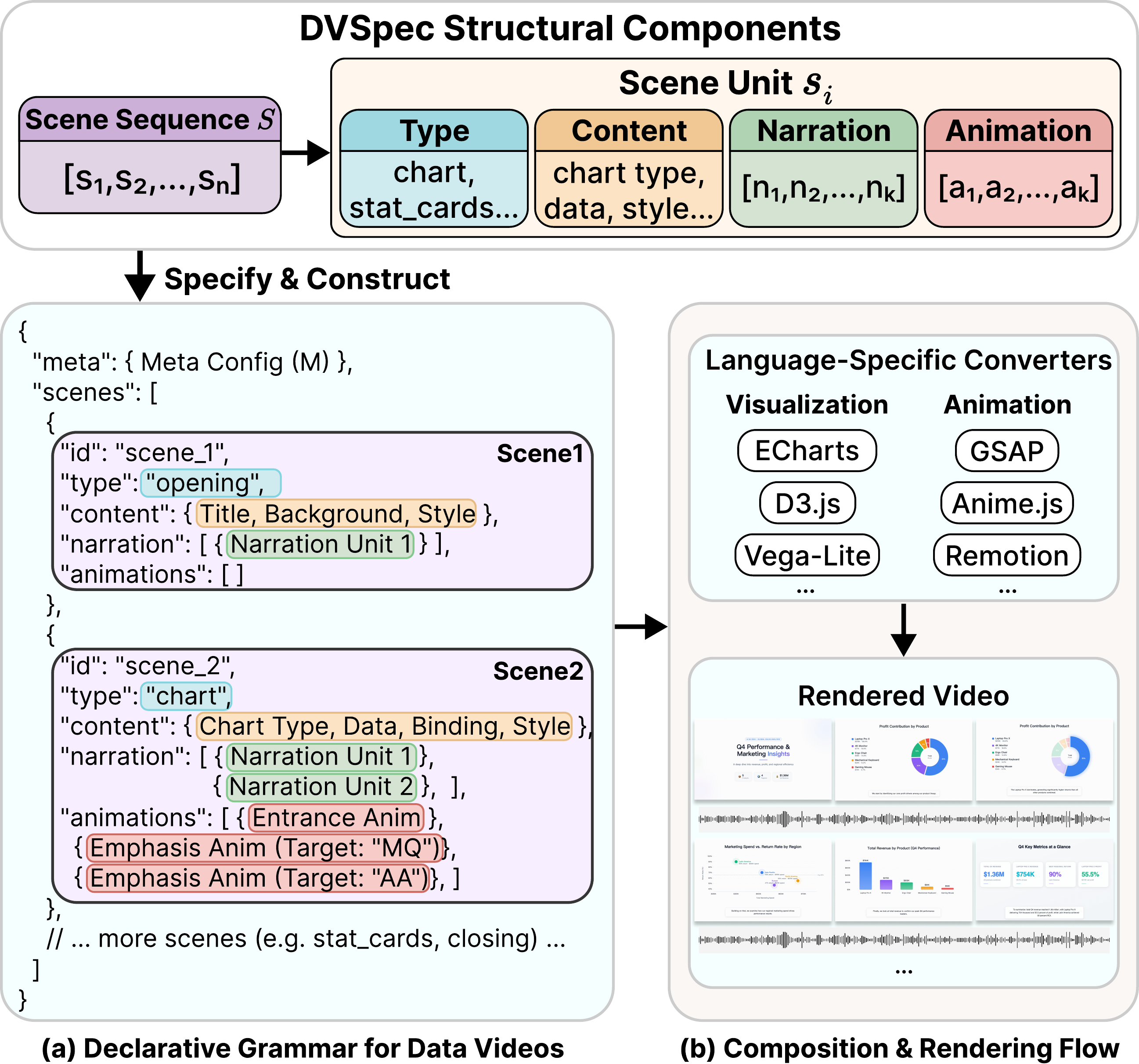}
  \caption{\dvspec structure and rendering flow.
  }
  \label{fig:dvspec}
\end{figure}

As shown in Figure~\ref{fig:dvspec}(a), \dvspec formalizes a data video as a combination of metadata $M$ and an ordered scene sequence $S = \langle s_1, \ldots, s_n \rangle$: $V := (M, S)$. Each scene $s_i$ is defined as a four-tuple $s_i := (\mathit{type},\allowbreak \mathit{content},\allowbreak \mathit{narration},\allowbreak \mathit{animation})$, where $\mathit{content}$ encapsulates visualization configuration (chart type, data bindings, style parameters), $\mathit{narration}$ is an ordered list of narration segments, and $\mathit{animation}$ is a list of animation effects. The scene-based design stems from research on data video narrative structures~\cite{amini2015understanding, shen2024dataplayer}.

\begin{figure*}[t]
  \centering
  \includegraphics[width=0.95\textwidth]{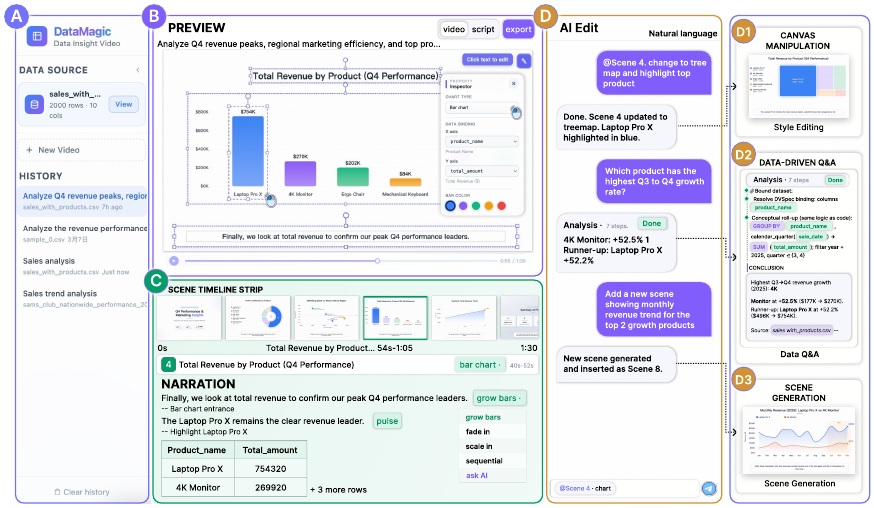}
  \caption{\system web interface with the main UI components (A-D), (D1-D3) highlight key interaction flows.
  }
  \label{fig:ui}
\end{figure*}


\dvspec introduces two key mechanisms to ensure data fidelity and audio-video synchronization:

\noindent\textbf{(1) Data-driven semantic references.} Visual elements are referenced through data attribute values (e.g., \texttt{\{"company": "Nvidia"\}}) rather than hard-coded identifiers, ensuring that references remain valid when data is updated or chart types change, and allowing every visual element to be precisely traced back to the underlying tabular data. Minor naming inconsistencies (e.g., casing or whitespace differences) are resolved through fuzzy matching at render time, and the pipeline applies a retry policy on code-execution failures.

\noindent\textbf{(2) Narration-index declarative triggering.} Animation trigger timing is declared using narration segment indices rather than absolute timestamps.
During rendering, the system automatically aligns animations based on the actual audio duration generated by text-to-speech (TTS), so that when users modify narration text, animation synchronization relationships are automatically maintained without manual keyframe adjustment~\cite{shen2024dataplayer}.

\subsection{Multi-Agent Generation Pipeline}
\label{sec:pipeline}

Data video generation involves complex dependencies among data processing, visual design, and narrative logic. Single-stage approaches struggle to ensure both per-scene data accuracy and overall narrative coherence. \system adopts a Generate-then-Orchestrate two-stage architecture, where parallel generation of candidate scenes enables the orchestration stage to perform global optimization rather than making greedy step-by-step decisions:

\stitle{Stage 1: Data-driven candidate generation.}
The Story Planner decomposes the user's high-level query into several independent analytical sub-tasks based on the analytical dimensions involved (e.g., temporal trends, category comparisons, regional distributions). For each sub-task, the Data Manager plans the data processing workflow and automatically generates Python code to extract, filter, and aggregate relevant data slices from the source table. The Visual Designer then designs the visualization scheme (chart type, data bindings, and style configuration), extracts key insights, and populates the content field of \dvspec. This stage generates a pool of candidate scenes in parallel, decoupling low-level data processing from high-level narrative construction.

\stitle{Stage 2: Global narrative orchestration.}
The Narration Director selects scenes from the candidate pool based on insight value and query coverage, and plans a playback order following narrative logic (e.g., ``macro-to-micro'', ``phenomenon before cause''). It generates narration for each scene conditioned on adjacent scenes' context, ensuring narrative coherence across scene transitions. The Animation Coordinator uses \dvspec's semantic reference mechanism to bind data entities mentioned in the narration to corresponding visual elements, achieving end-to-end audio-video synchronization.

The generated \dvspec is compiled into the final video by the rendering engine, with narration text synthesized into speech via TTS. The system adopts a model-agnostic design to support different LLM backends.

\subsection{Provenance-Based Interaction and Exploration}
\label{sec:editing}

Prior work has established data provenance as a key enabler for interactive visualizations~\cite{psallidas2018provenance}. \system extends this idea to the data video setting: \dvspec records the binding relationships from visual elements to underlying data fields during the generation stage. This structured data provenance information permeates the system's downstream interactions, supporting two core capabilities.

\stitle{Parametric Editing.}
Since \dvspec explicitly records the data fields and narration segments corresponding to each visual and animation element, any modification by the user can be precisely located and applied locally, without regenerating the entire video~\cite{tang2026sketch, shen2026debugging}.
The system provides three interaction modes: canvas direct manipulation (e.g., switching chart types, adjusting data mappings), declarative script editing (directly modifying narration or animation parameters), and natural language commands (e.g., ``@Scene 4: change to treemap and highlight top product''). All three share the same \dvspec state, maintaining context consistency across mode switches.

\stitle{Data Q\&A.}
The provenance relationships preserved by \dvspec enable the system to map users' natural language questions to specific data binding fields: the system uses an LLM to parse the question~\cite{shen2022towards, luo2021natural, wu2024chartinsights}, identifies relevant data fields from the current scene's \dvspec bindings, and constructs structured query operations (e.g., filtering, aggregation, extremum retrieval) to execute directly on the underlying tabular data~\cite{li2024dawn}, rather than relying on visual models to infer content from pixels. Users can also transform insights discovered through Q\&A into new video scenes, completing the full loop from data exploration to narrative expansion.


\section{Demonstration Scenarios}
\label{sec:demo}


We design three scenarios to showcase \system's capabilities in end-to-end data video generation, multimodal editing, and provenance-based data Q\&A,  with quantitative evaluation to validate generation quality. The demonstration is conducted through a web interface (Figure~\ref{fig:ui}), which includes a data and history panel (A), a real-time video preview panel (B), a scene timeline and narration editor (C), and an AI-assisted editing panel (D). Users interact with the system directly at each stage of the workflow.

\subsection{Scenario 1: Data-Driven Automated Generation}
\label{sec:scenario1}

This scenario demonstrates how \system transforms raw data into a narrative video.
%
{The user uploads a business CSV file (e.g., technology company financial reports) and enters a query (e.g., ``Analyze Q4 revenue peaks, regional marketing efficiency, and top product profit contributors'').}
The user can observe the multi-agent engine executing in real time: query decomposition, data slicing, chart design, and narrative orchestration.

Once generation is complete, the video plays in the embedded player (Figure~\ref{fig:ui} B). Unlike traditional black-box generation, the system transparently displays the corresponding scene timeline, narration text, animation labels, and underlying data tables in region C, allowing the audience to intuitively understand the strict binding between visual content and underlying data fields.
The history panel reveals intermediate outputs for traceable generation.

\subsection{\dvspec{}-Based Multimodal Editing}
\label{sec:scenario2}

Data reporting is inherently iterative. Traditional pixel-level videos are extremely difficult to modify after generation, whereas \system enables parametric editing through \dvspec. The three interaction modes share the same \dvspec state, and the audience can freely switch between them based on the video generated in Scenario~1:

\noindent\textbf{(1) Canvas direct manipulation.} 
%
{The audience clicks to select a chart on the video canvas and the system displays a Property Inspector panel.}
Users can switch chart types via dropdown menus (e.g., convert bar to pie chart), modify data field mappings, or adjust color schemes, as well as drag elements or double-click to edit text. All interactions are mapped in real-time to local \dvspec updates.

\noindent\textbf{(2) Declarative script editing.} The audience directly modifies narration text or chart parameters in the narration editor (region C). When text changes cause audio duration to change, the system automatically re-aligns animation trigger timing without manual keyframe adjustment.

\noindent\textbf{(3) Natural language commands.} The audience enters commands in the AI chat panel, e.g., ``@Scene 4: change to treemap and highlight top product'' (as shown in Figure~\ref{fig:ui} D1). The system parses the command and automatically updates the corresponding scene's \dvspec configuration, triggering real-time incremental re-rendering.

\subsection{Scenario 3: Structured Provenance-Based Data Q\&A}
\label{sec:scenario3}

This scenario demonstrates how \system transforms one-way data videos into explorable interactive data interfaces. Traditional video models output pure pixel streams and cannot understand the data meaning behind the visuals, whereas \dvspec preserves the complete semantic and data context, enabling precise structured retrieval and data provenance rather than relying on visual models to infer content from pixels.

The audience asks questions about the currently playing video scene (e.g., ``Which product has the highest Q3 to Q4 growth rate?''). As shown in Figure~\ref{fig:ui} D2, the system locates the data binding fields for the current scene in \dvspec, jointly queries the original dataset, and returns an accurate answer (e.g., 4K Monitor: +52.5\%, Runner-up: Laptop Pro X +52.2\%~$\uparrow$). The audience can also ask global questions, experiencing multi-granularity data exploration from scene-level to video-level.

Furthermore, the audience can instruct the system to transform Q\&A insights into new video scenes (e.g., ``Add a new scene showing monthly revenue trend for the top 2 growth products''). As shown in Figure~\ref{fig:ui} D3, the system automatically generates the corresponding \dvspec configuration and inserts the new scene into the video sequence, completing the full loop from data exploration to narrative expansion.

\subsection{Quantitative Evaluation}
\label{sec:eval}


We evaluated \system on 109 real-world samples from DAComp-DA~\cite{lei2025dacomp} and T2R-bench~\cite{zhang2025t2r}.
We first used four LLMs (DeepSeek-V3.2~\cite{deepseek2025v3}, Gemini-2.5-Pro~\cite{google2025gemini25pro}, GPT-5~\cite{openai2025gpt5}, and Claude-Sonnet-4~\cite{anthropic2025claudesonnet4}) to directly generate data videos in one pass, probing current capability limits;
we also integrated these models into the \system framework to verify model-agnosticism.
The evaluation covered execution rate (the fraction of samples for which the full pipeline completes without error and renders a playable video) and five quality dimensions (Intent Fulfillment, Data Insights, Narrative Quality, Animation Effectiveness, Aesthetic Quality; 1--5 scale)~\cite{xie2025visjudgebench, tang2026igenbench, su2026vcg, luo2021synthesizing, luo2026nvbench}.
Gemini-2.5-Pro~\cite{google2025gemini25pro} served as an automated judge (Pearson $r{=}0.91$ vs.\ human experts on 60 sampled videos).


Results showed that even the strongest LLMs faced significant challenges in data video generation: 
{average quality scores ranged from 1.91 to 2.22, with execution rates between 48--86\%. }
The main issues were audio-visual temporal misalignment and lack of narrative structure: 
models frequently produced videos where narration was out of sync with visuals and scenes lacked logical transitions. 
\system addressed these problems through \dvspec's narration-index triggering mechanism 
and the Generate-then-Orchestrate strategy, raising the average score to 3.89 with execution rates above 95\%. 
The largest gains were in Animation Effectiveness ($1.84{\rightarrow}4.03$, +120\%)
and Narrative Quality ($2.00{\rightarrow}3.43$, +72\%).
On average, direct generation completes in ${\sim}57$ seconds; \system totals ${\sim}176$ seconds (60\,s configuration generation $+$ 116\,s rendering at parallelism\,=\,3), reflecting a deliberate quality--latency trade-off.
Ablation experiments confirm that both components are indispensable: removing the Story Planner degrades average quality by 11.6\% (Narrative $-$15.1\%), and removing Orchestration by 9.0\% (Intent $-$12.4\%).
Training a dedicated data video model could further reduce API dependency and latency~\cite{lin2025lead, qi2026pariskv}.

\begin{acks}
This paper was supported by the NSF of China (62402409); Youth S\&T Talent Support Programme of Guangdong Provincial Association for Science and Technology (SKXRC2025461); the Young Talent Support Project of Guangzhou Association for Science and Technology (QT-2025-001); Guangzhou Basic and Applied Basic Research Foundation (2026A1515010269, 2025A04J3935, 2023A1515110545); Guangzhou-HKUST(GZ) Joint Funding Program (2025A03J3714); and Research and Development of Heterogeneous Computing Interconnection and Scheduling Software Stack (YF202400000003).
\end{acks}


\bibliographystyle{ACM-Reference-Format}
\vspace{-0.5em}
\bibliography{reference}

\end{document}